# Long-Lived Waveguides and Sound Wave Generation by Laser Filamentation

Liad Levi[*], Oren Lahav[*], Ron A. Nemirovsky, Jonathan Nemirovsky, Itai Orr, Ido Kaminer, Mordechai Segev and Oren Cohen

*Solid state institute and physics department, Technion, Haifa, Israel 32000*

We discover long-lived (microsecond-scale) optical waveguiding in the wake of atmospheric laser filaments. We also observe the formation and then outward propagation of the consequent sound wave. These effects may be used for remote induction of atmospheric long-lived optical structures from afar which could serve for a variety of applications.

[*] These authors contributed equally to this work.

Propagation of self-guided laser filaments through air and other gases results in a rich variety of phenomena and applications [1-3]. A laser filament is formed when a femtosecond pulse, with peak intensity above the critical power for collapse (3 GW in air at an optical wavelength of 800 nm), is propagating in a transparent medium [1]. In air, the diameter of a filament is approximately 100 µm and it can propagate over distances much longer than the Rayleigh length, from 10 cm up to the kilo-meter range [4-7]. A filament is formed due to a dynamic balance between the linear diffractive and dispersive properties of the medium and its nonlinear features such as self-focusing optical Kerr effect and defocusing due to the free electrons which are released from molecules through multi-photon ionization. In the atmosphere, filaments can be initiated at predefined remote distances [4,5] and propagate through fog, clouds and turbulence [8,9]. Thus, filaments are attractive for atmospheric applications such as remote spectroscopy [8,10] and laser-induced water condensation [11].

An atmospheric filament pulse initiates complex nonlinear dynamics in the densities of free electrons and ions, air density, and in the level of molecular alignment. However, all of these effects are generally believed to die out after ~10 nanoseconds. Experimental data about the long term dynamics is especially scarce. The filamenting pulse leaves behind free electrons at initial densities of $10^{16}$-$10^{17}$ cm$^{-3}$ [2,3], mostly from multi-photon ionization of oxygen molecules because the ionization potential of $N_2$ molecules is significantly larger (12 eV and 16 eV for $O_2$ and $N_2$, respectively). Initially, the free electron density exhibits a radial bell-shape profile. It is now established that 100 picoseconds to ~ 1 nanosecond later, a shock wave of electron density forms and propagates outward supersonically [12-15]. The resultant radially increasing plasma density can be used for guiding a delayed laser pulse [12-15] or a microwave pulse [16]. However, recombination between the free electrons and positive oxygen molecules decreases the plasma density by two orders of magnitude and limits this waveguiding effect to a few nanoseconds. Such nanosecond dynamics of charge carriers also results in the emission of a high power burst of terahertz radiation [17]. Subsequently, when the density of free electrons decreases to below $10^{14}$ cm$^{-3}$, the capture of free electrons by neutral oxygen molecules (with characteristic time of 150 nanoseconds), becomes dominant [5,18]. The influence of this population on the refractive index has not yet been explored, thus it is yet unknown if it can contribute to waveguiding or anti-guiding effects. Another effect induced by the filamentation is impulsive rotational excitation in the $N_2$ and $O_2$ molecules [19], where the molecules exhibit periodic revivals of molecular alignment for several tens of picoseconds after the passing of the filament pulse. Consequently, the refractive index of air undergoes anisotropic temporal modulation [20]. This mechanism was used for guiding properly delayed picosecond pulses [21,22], but at times larger than several nanosecond after the filamenting pulse, even this process does not leave behind any waveguiding effects. ***In fact, all processes resulting from plasma or molecular alignment in the wake of atmospheric laser filaments are limited to the first few nanoseconds period.*** Consequently, it was generally believed that ten nanoseconds after the filament, the medium does not exhibit any waveguiding effect. In contrast to that, it was recently discovered that 0.1-1 milliseconds after the filament, there is a circular negative index change that acts as an antiguide by defocusing a probe beam [23]. This effect was attributed to reduction in the air density at the center of the filament as a result of heating. Altogether, to the best of our knowledge, thus far all experiments and theories on laser filamentation in the atmosphere concluded that there is no long-lived (>10 nanosecond) waveguiding effect left behind the femtosecond filamenting pulse. This severely limits any CW application of laser filamentation, because the repetition rate of any high power laser used for creating the filament is low, hence for most of the time between pulses light would not be guided. Likewise, any other potential application would have to "live" on a picosecond scale, because at later times, waveguiding by the filament was thought to be nonexistent.

Here, we show the exact opposite: we demonstrate theoretically and experimentally that the filament induces a transient positive index change which lasts for approximately 1.5 microseconds. Through numerical simulations, we show that the positive index change results from a transient increase in the air density. We demonstrate waveguiding through this induced positive index change. In addition, we observe and study the formation of ultra-short acoustic waves and subsequently their outward propagation at the speed of sound. These induced long-lived waveguides can be useful for numerous applications of laser filemantation in the atmosphere, from power transmission through these channels to backward propagation of coherent and incoherent radiation for remote sensing.

Our experimental methodology relies on using an electronically-delayed short optical pulse for probing the long-term effects generated by a femtosecond filementing pulse in the atmosphere. The experimental setup is shown in Fig. 1. A "pump" Ti:Sapphire (wavelength centered around 0.8 µm) pulsed laser beam 1 cm wide with 50 fs time-duration, 1 mJ energy/pulse and at 1 KHz repetition rate is focused to a diameter of ~100 μm using a f=100 cm lens. The expected Raleigh range of such a beam is of the order of 2 cm. Our beam forms a filament of ~30 cm length in the free air. We probe the filament wake using a weak pulsed laser beam of wavelength 527 nm, with 150 ns pulse-duration and 1 KHz repletion rate, which is triggered by the femtosecond laser. The delay between the probe pulse and the pump pulse is controlled electronically and can span a range of one msec. The pump and probe beams propagate in opposite directions. This setting helps in unraveling the induced index change, as the phase change accumulates along the filament, and at the same time the counter-propagation geometry makes it easy to separate the probe beam from the pump beam. We define the "input" and "output" planes as the entrance and exit planes of the *probe* pulse, which is propagating within the channel induced by the filament. A lens, a movable 4f system and a CCD camera are used to image the probe beam at the input and output planes.

In the first experiment, we expand the probe beam such that it is approximately a plane wave at the input. Figures 2(a-f) show the intensity of the output probe beam for several delay times with respect to the filament pulse (where Δt=0 corresponds to time delay at which the centers of the probe and pump pulses collide in the filament region). The ratio between the peak intensity at the center ($I_0$) and the intensity in the uniform region ($I_{BG}$), as a function of the time delay, is shown in Fig. 2g. During the first 2/3 microseconds, the intensity of the output probe beam in the region of the center of the filament wake increases. The period within which the intensity at the center is larger than the background lasts for approximately 1.5 microseconds. This intensity profile suggests the presence of a positive index change (waveguiding effect) at the center of the filament wake, which "pulls" the light into the center. We test the sensitivity of this central waveguiding effect to wind (artificially generated by an air blow gun) and observe that the positive index effect is robust even under strong wind. Moreover, we find that the positive index is induced in a broad range of experimental parameters. As an example, Fig. 2g shows the increased peak intensity effect for several values of the pump pulse energy.

To test the presence of the central positive refractive index change induced by the femtosecond filament, we check if the filament wake can guide our probe pulse. In this experiment, we insert a f=100 cm lens that focuses the probe beam into a diameter of ~100 μm at the input plane (Fig. 3a). When the pump beam is blocked, hence no filament is formed, the probe beam is propagating in free air and is broadening considerably due to diffraction (Fig. 3b). In contrast, after a filament has formed, the probe beam is well guided within the positive index change left behind. For example, Fig. 3c shows the intensity of the probe beam at the output plane 0.8 μs after the pump pulse has created the 30 cm filament, clearly, showing that the beam is indeed guided within the waveguide induced by the filament. Finally, Fig. 3d shows the fraction of power localized within the guiding region (a circle with diameter of 300 microns) as a function of the time delay. This experiment unequivocally shows that the filament induces a positive index structure (waveguiding effect) surviving ~1.5 microseconds after the 50 fsec filament pulse has passed, and that such waveguiding effect can be used for long-term guiding of another beam.

In order to understand the physical mechanism that gives rise to the positive index change, we simulate the dynamics of the air density using hydrodynamic simulations, following the model presented in [23] with the addition of radiative cooling – a known important effect in thin flames [24, 25]. The simulated dynamics is depicted in Fig. 4a. As shown there, shortly after the pump pulse has initiated the filamentation process, the air expands outwards from the beam axis. This happens because the pump beam heats the gas and therefore creates a local pressure peak. During the outward expansion, the hot gas cools down by radiation, along with mechanical work and heat conduction. The combination of density decrease together with the cooling reverses the pressure gradient and leads to inflow of air towards the center, and to the formation of central region whose density is higher than the background air density (Fig. 4a and inset therein). The air spike, which is also manifested in a considerable increase in the refractive index – as shown in our experiments presented in Figs. 3c and 3d – is most pronounced ~500 nanoseconds after the pump pulse. At later times, the air spike relaxes into an expanding ring of higher density air. Interestingly, this dynamics is somewhat analogous to the formation of a Worthington jet and subsequent surface wave as a result of an object impact into a still liquid surface [26]. Figures 4b and 4c show the simulated intensity patterns of an initially uniform beam propagating through 30cm of air. More specifically, these figures present the calculated intensity patterns going through the perturbed air-density profiles, at 500 and 1100 nanoseconds delay times, respectively. The agreement with the experimental images (Figs. 2c and 2e) is significant.

Next, we present the formation and outgoing-propagation of a radial sound-wave pulse produced in the wake of the filament. Notably, the formation of a sound wave from laser filaments was studied using only acoustical probing [27]. The acoustic wave is formed during the same time-window as the central spike with the positive index change. The actual formation of the leading ring crest of the acoustic pulse is shown in Figs. 2(a-f) and also in Fig. 5a which shows line cuts of Figs. 2(d,e,f). Figure 5b shows the radius of the leading crest as a function of $\Delta t$. Interestingly, the leading crest is formed at 0.08 µsec, yet it starts to propagate outwards only at 0.3 µsec delay times. The second and third crest rings are formed at 0.78 µsec and 0.62 µsec delay times, respectively. Finally, Fig. 6 shows the outgoing propagation of the sound wave. Figures 6(a-e) show the intensity of the probe beam (for a plane-wave launch configuration similar to that of Fig. 2) at the output plane for several delay times. These plots clearly show the three-cycle acoustic pulse. Figure 6f shows the radius of the leading crest versus delay time from which we calculated the wave velocity to be 333±1 m/s. This value is comparable to the sound velocity in air, thereby showing that the wave is indeed an acoustic wave. After the radial acoustic wave is emitted, a negative index change is left behind at the center. This negative index change, which decays slowly and is still observed even after a millisecond, was explored in Ref. 23. However, the temporal resolution (~40 microseconds) in the experiment of [23] was set by an electronic shutter that controlled the input light to the CCD. As such, that experiment was insensitive to the dynamics during the first several microseconds after the passing of the filamenting pulse. For this reason, the experiments in [23] could not reveal the waveguiding effects presented here.

In conclusion, we discovered that a filament induces a transient positive index-change that lasts for approximately 1.5 microsecond and demonstrated waveguiding through it. Detailed hydrodynamic simulations with radiative cooling show that this positive index change is the result of a transient air density peak occurring along the filament axis. We also explored the formation and propagation of the sound wave that results from the filament. To our knowledge, this is the first experiment demonstrating that femtosecond laser filementation in the atmosphere leaves behind it a long lasting (microseconds) waveguiding channels. With combination of current fiber laser technology that can drive filaments at repetition rate that approaches MHz scale [28,29], a CW laser beam can already be almost continuously guided. This ability to guide probe beams for such relatively long times offers many applications. For example, power channeling of light over large distances [30]. Naturally, the ability to guide light over microsecond (instead of sub-picosecond, as was the common belief thus far) improves the potential efficiency of power transmission by orders of magnitude. In a similar vein, laser filamentation can give rise to lasing [31,32]. Clearly, having waveguiding effects for extended periods of time would greatly improve applications of these lasing effects. Finally, filaments can be used for writing refractive index gratings at preselected positions in the atmosphere [33] that may be used for enhancing the stability of following filaments [34,35] and for increasing the numerical aperture of telescopes [36]. Clearly, having long-lived index changes induced by the filaments introduces many new applications, which are yet to be explored.

**Figures**

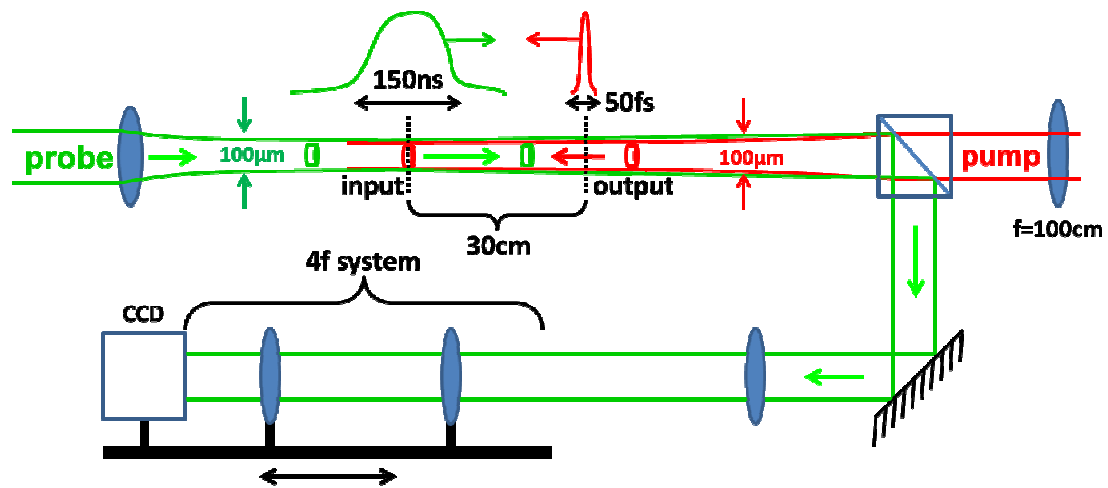

Figure 1: (Color online) Scheme of the experimental setup. A Ti:Sapphire pulsed laser 1 cm wide beam with 50 fs time duration, central wavelength of 800 nm, 1 mJ energy/pulse at 1KHz repetition rate, is focused into a diameter of ~100 μm, creating a filament of ~30 cm in the free air. The probe beam is a weak 527 nm pulse laser beam, with a duration time of 150 ns and repetition rate of 1KHz, triggered by the femtosecond laser. The delay between the pulses is controlled electronically and can span a range of one msec. In the filament region, the probe and pump beams propagate in opposite directions. The "input" and "output" planes correspond to the *probe* pulse entrance and exit planes of the filament channel, respectively. An imaging system images the probe pulse at the input and output planes. In the experiment presented in Fig. 3, we insert a lens that focuses the probe beam to a diameter of ~100 μm at the input plane. No focusing lens is used in the experiments presented in Figs. 2 and 4, hence the probe beam is approximately a plane wave.

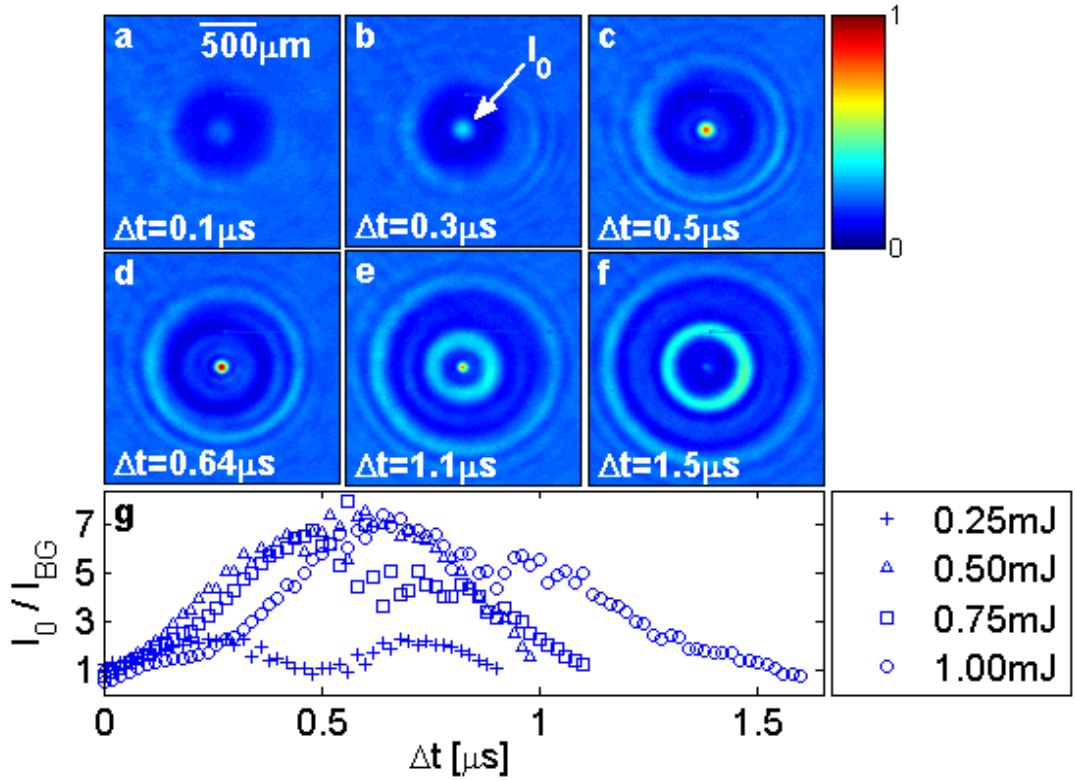

Figure 2: (Color online) Long-lived positive index change in the filament wake. Shown are the intensity patterns of the probe beam at the output plane, for delay times of $\Delta t=0.1$ (a), $\Delta t=0.3$ (b), $\Delta t=0.5$ (c), $\Delta t=0.64$ (d), $\Delta t=1.1$ (e) and $\Delta t=1.5$ (f) microseconds after the filamenting pulse. (g) Ratio between the peak intensity at the center ($I_0$), which is pointed by an arrow in (b), and the background intensity ($I_{BG}$) versus the delay time between the probe and pump pulses for several pulse energies. Here, $\Delta t=0$ corresponds to time delay for which the probe and pump pulses collide within the filament channel. The outer ring that forms in (a) and subsequently expands and becomes stronger (b-f) reflects an acoustic density wave discussed in Figs. 5 and 6.

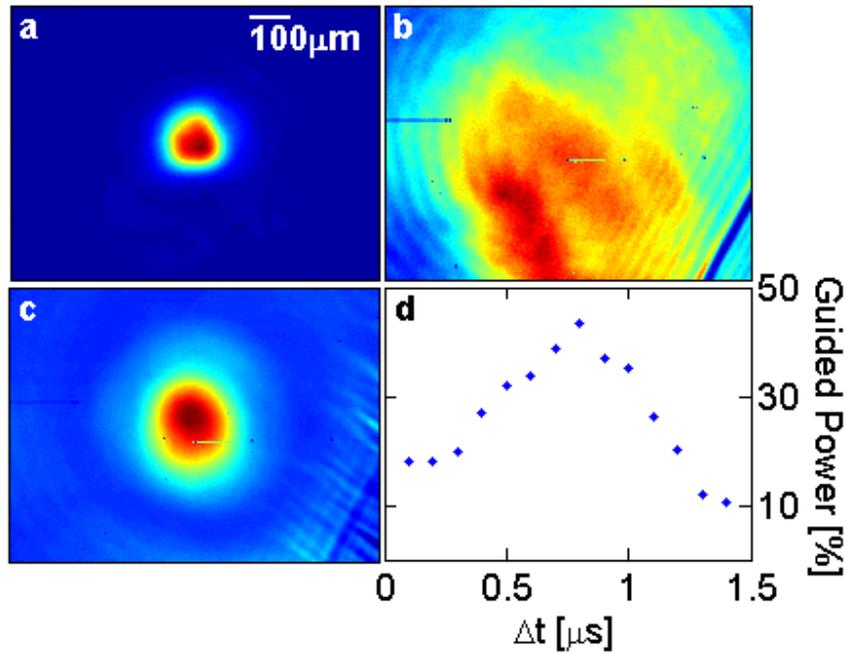

Figure 3: (Color online) Long-lived filament-induced waveguide. (a) Intensity structure of the focused probe beam at the input plane. (b) Intensity structure of the probe beam at the output plane when the pump beam is blocked, i.e. no filament is formed. In the absence of the filament, the probe beam exhibits considerable diffraction broadening. (c) Intensity structure of the guided probe beam at the output plane for delay time of 0.8 µs with respect to the filamenting pulse. (d) The fraction of power localized within the guiding region (a circle with 300 µm diameter). In Figs 3(a-c), each plot is normalized separately. The line in the lower right corners of Figs 3(b,c) corresponds to a wire that we used for spotting the output plane.

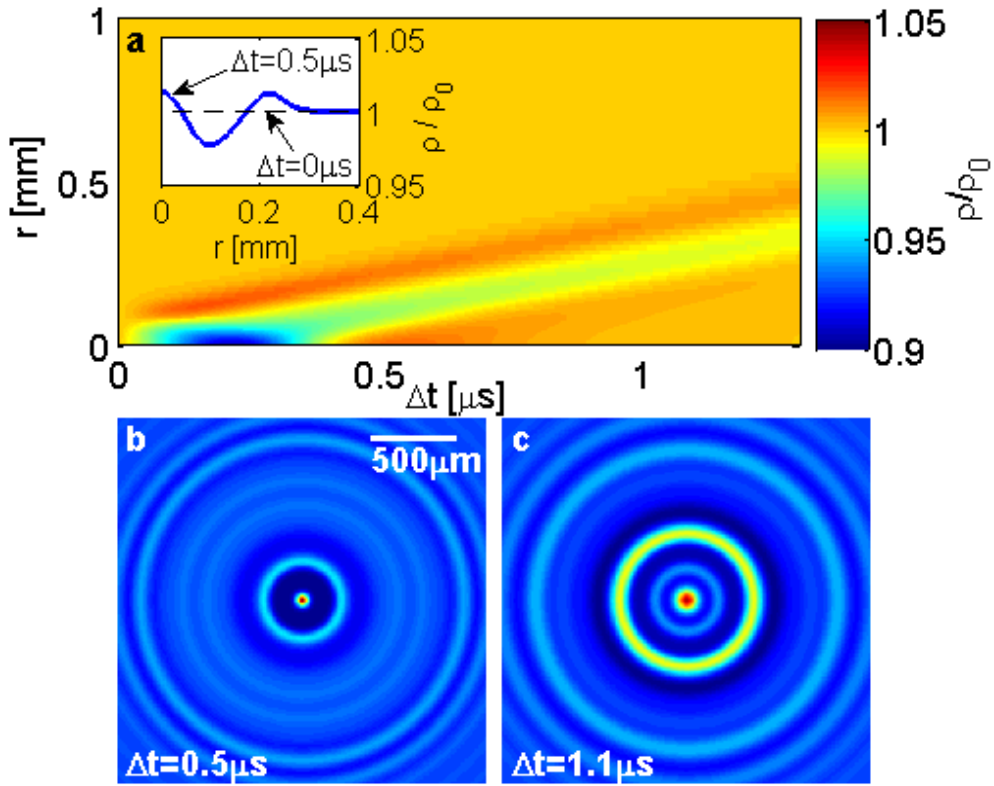

Figure 4: (Color online) Numerical simulations of the filament wake. (a) Radial air density ($\rho(r)$) as a function of time delay. At the beginning ($\Delta t=0$) the medium is homogeneous, i.e. $\rho(r)=\rho_0$. The density at the center (r = 0) initially becomes lower due to the sudden heating. Subsequently, the temperature is lowered by rapid radiative cooling and air starts to move inwards, leading to increased air density at the center. Inset: density profiles at time delays of $\Delta t=0$ μsec (dashed line) and $\Delta t=0.5$ μsec (solid line) after the filamenting pulse. (b), (c) Simulation results showing the optical intensity patterns of the probe beam at the output plane after propagating through 30cm of air, at delay times of 0.5 and 1.1 μsec, respectively.

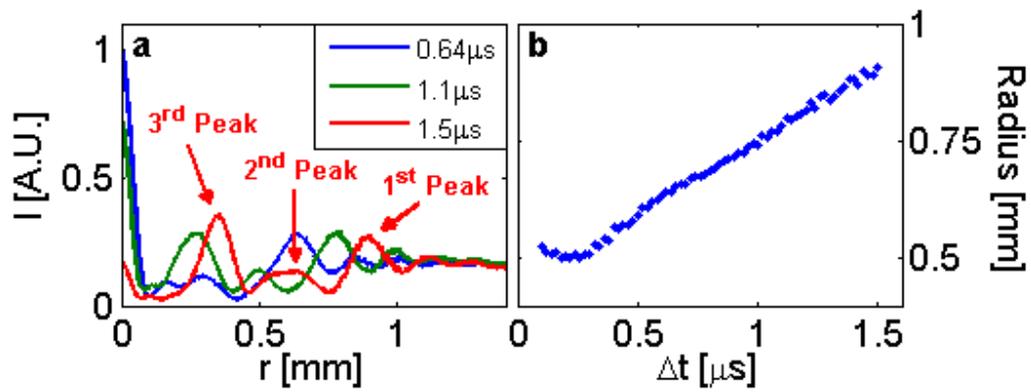

Figure 5: (Color online) Formation of the acoustic pulse. (a) 1D Intensity profiles of the probe beam at the output plane at 0.64, 1.1 and 1.5 microseconds delay times. (b) Radius of the first ring crest (1st peak) as a function of time delay.

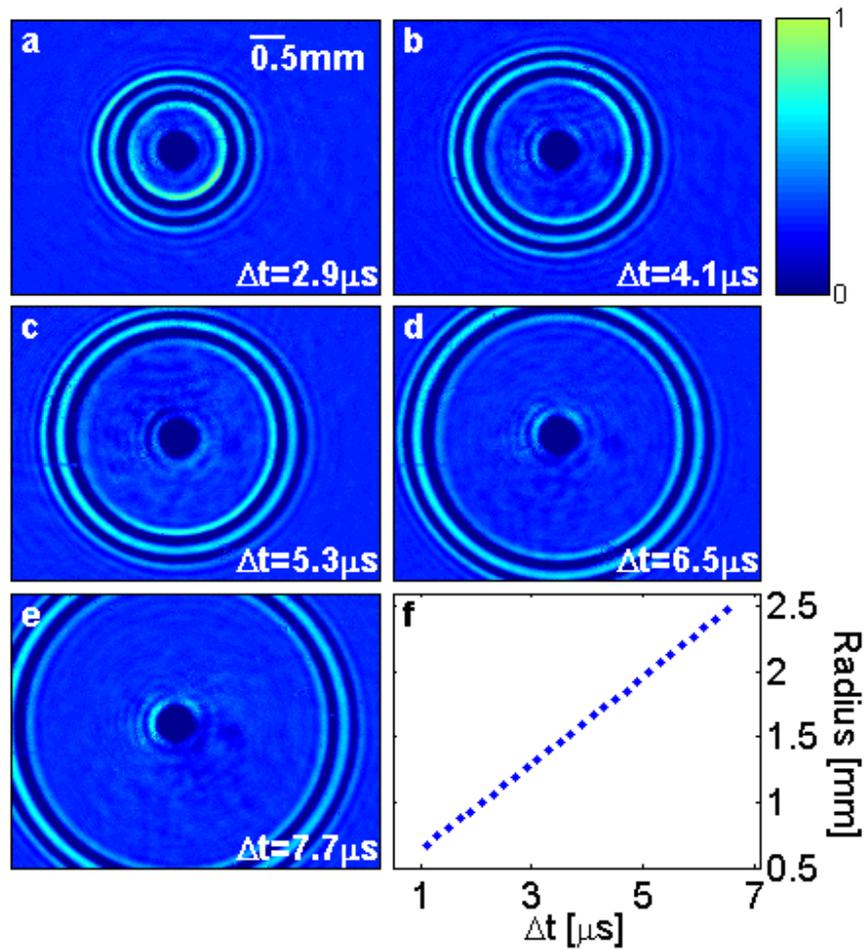

Figure 6: (Color online) Outward propagation of the acoustic pulse. Shown are the intensity structures of the probe beam (plane wave) at the output plane for delay times of Δt=2.9 (a), Δt=4.1 (b), Δt=5.3 (c) Δt=6.5 (d) and Δt=7.7 (e) microseconds after the filamenting pulse. (f) Radius of the leading crest versus delay time. The calculated angle of the line corresponds to a velocity of 333 ± 1 m/s.